\documentstyle[12pt]{article}
\newcommand{\dd}{{\rm d\kern-.17em I}}
\newcommand{\w}{\wedge \kern-.8em \wedge}

\newcommand{\real}{{\rm I\kern-.17em R}}

\newcommand{\au}{
\begin{picture}(11,3)(-2,-2)
\put(5,8){$\scriptscriptstyle{1}$} \put(0,-2){$A$}
\end{picture}}

\newcommand{\ad}{
\begin{picture}(11,3)(-2,-2)
\put(5,8){$\scriptscriptstyle{2}$} \put(0,-2){$A$}
\end{picture}}

\newcommand{\bu}{
\begin{picture}(11,3)(-2,-2)
\put(4,8){$\scriptscriptstyle{1}$} \put(0,-2){$B$}
\end{picture}}

\newcommand{\bd}{
\begin{picture}(11,3)(-2,-2)
\put(4,8){$\scriptscriptstyle{2}$} \put(0,-2){$B$}
\end{picture}}

\newcommand{\tbu}{
\begin{picture}(11,3)(-2,-2)
\put(3,11){$\scriptscriptstyle{1}$} \put(-1,-2){${\tilde B}$}
\end{picture}}

\newcommand{\tbd}{
\begin{picture}(11,3)(-2,-2)
\put(3,11){$\scriptscriptstyle{2}$} \put(-1,-2){${\tilde B}$}
\end{picture}}

\newcommand{\fu}{
\begin{picture}(11,3)(-2,-2)
\put(3,8){$\scriptscriptstyle{1}$} \put(-1,-2){$F$}
\end{picture}}

\newcommand{\fd}{
\begin{picture}(11,3)(-2,-2)
\put(3,8){$\scriptscriptstyle{2}$} \put(-1,-2){$F$}
\end{picture}}

\newcommand{\tfu}{
\begin{picture}(11,3)(-2,-2)
\put(3,11){$\scriptscriptstyle{1}$} \put(-1,-2){${\tilde F}$}
\end{picture}}

\newcommand{\tfd}{
\begin{picture}(11,3)(-2,-2)
\put(3,11){$\scriptscriptstyle{2}$} \put(-1,-2){${\tilde F}$}
\end{picture}}

\newcommand{\fiu}{
\begin{picture}(11,3)(-2,-2)
\put(3,6){$\scriptscriptstyle{1}$} \put(-1,-3){$\varphi$}
\end{picture}}

\newcommand{\fid}{
\begin{picture}(11,3)(-2,-2)
\put(3,6){$\scriptscriptstyle{2}$} \put(-1,-3){$\varphi$}
\end{picture}}

\newcommand{\piu}{
\begin{picture}(9,3)(-2,-2)
\put(1,6){$\scriptscriptstyle{1}$} \put(-1,-3){${\tilde \pi}$}
\end{picture}}

\newcommand{\pid}{
\begin{picture}(9,3)(-2,-2)
\put(1,6){$\scriptscriptstyle{2}$} \put(-1,-3){${\tilde \pi}$}
\end{picture}}

\newcommand{\nuu}{
\begin{picture}(9,3)(-2,-2)
\put(1,6){$\scriptscriptstyle{1}$} \put(-1,-3){${\tilde \nu}$}
\end{picture}}

\newcommand{\nud}{
\begin{picture}(9,3)(-2,-2)
\put(1,6){$\scriptscriptstyle{2}$} \put(-1,-3){${\tilde \nu}$}
\end{picture}}

\newcommand{\psiu}{
\begin{picture}(11,3)(-2,-2)
\put(3,9){$\scriptscriptstyle{1}$} \put(-1,-3){${\tilde \psi}$}
\end{picture}}

\newcommand{\psid}{
\begin{picture}(11,3)(-2,-2)
\put(3,9){$\scriptscriptstyle{2}$} \put(-1,-3){${\tilde \psi}$}
\end{picture}}

\newcommand{\sigu}{
\begin{picture}(9,3)(-2,-2)
\put(1,6){$\scriptscriptstyle{1}$} \put(-1,-3){$\sigma$}
\end{picture}}

\newcommand{\sigd}{
\begin{picture}(9,3)(-2,-2)
\put(1,6){$\scriptscriptstyle{2}$} \put(-1,-3){$\sigma$}
\end{picture}}

\newcommand{\tmu}{
\begin{picture}(11,3)(-2,-2)
\put(3,6){$\scriptscriptstyle{1}$} \put(-1,-3){${\tilde m}$}
\end{picture}}

\newcommand{\tmd}{
\begin{picture}(11,3)(-2,-2)
\put(3,6){$\scriptscriptstyle{2}$} \put(-1,-3){${\tilde m}$}
\end{picture}}

\newcommand{\cu}{
\begin{picture}(11,3)(0,0)
\put(4.6,2){$\scriptscriptstyle{1}$}
\put(0,-2){\large$\nabla$\normalsize}
\end{picture}}

\newcommand{\cd}{
\begin{picture}(11,3)(0,0)
\put(4.6,2){$\scriptscriptstyle{2}$} \put(0,-2){\large
$\nabla$\normalsize}
\end{picture}}

\newcommand{\etu}{
\begin{picture}(8,3)(0,0)
\put(1,7){$\scriptscriptstyle{1}$} \put(0,0){$\eta$}
\end{picture}}

\newcommand{\etd}{
\begin{picture}(8,3)(0,0)
\put(1,7){$\scriptscriptstyle{2}$} \put(0,0){$\eta$}
\end{picture}}

\newcommand{\but}{
\begin{picture}(9,3)(-1,-1)
\put(3,10){$\scriptscriptstyle{1}$}\put(0,0){$\beta$}
\end{picture}}

\newcommand{\bdt}{
\begin{picture}(9,3)(-1,-1)
\put(3,10){$\scriptscriptstyle{2}$}\put(0,0){$\beta$}
\end{picture}}

\newcommand{\lu}{
\begin{picture}(7,3)(-1,-1)
\put(0,10){$\scriptscriptstyle{1}$}\put(0,0){$l$}
\end{picture}}

\newcommand{\ld}{
\begin{picture}(7,3)(-1,-1)
\put(0,10){$\scriptscriptstyle{2}$}\put(0,0){$l$}
\end{picture}}

\newcommand{\nnu}{
\begin{picture}(9,3)(-1,-1)
\put(1,7){$\scriptscriptstyle{1}$}\put(0,0){$n$}
\end{picture}}

\newcommand{\nnd}{
\begin{picture}(9,3)(-1,-1)
\put(1,7){$\scriptscriptstyle{2}$}\put(0,0){$n$}
\end{picture}}

\newcommand{\su}{
\begin{picture}(7,3)(-1,-1)
\put(2,6){$\scriptscriptstyle{1}$}\put(0,0){$s$}
\end{picture}}

\newcommand{\sd}{
\begin{picture}(7,3)(-1,-1)
\put(2,6){$\scriptscriptstyle{2}$}\put(0,0){$s$}
\end{picture}}

\begin{document}

\baselineskip=22pt plus 0.2pt minus 0.2pt \lineskip=22pt plus
0.2pt minus 0.2pt
\begin{center}
\vspace*{1cm} \LARGE
BF Actions for the Husain-Kucha\v{r} Model.\\

\vspace*{1.5cm}

\large

J.\ Fernando\ Barbero\ G.\footnote{barbero@laeff.esa.es},\\and\\
Eduardo J. S. Villase\~nor\footnote{ejesus.sanchez@mat.ind.uem.es}

\vspace*{1.5cm}

\normalsize
{\it Escuela Superior de Ingenier\'{\i}a Industrial,\\
Universidad Europea \\
Urb. El Bosque, C/ Tajo s/n\\
Villaviciosa de Od\'on, Madrid, 28670\\
Spain\\\hspace{5mm}\\}

\vspace{.3in}
November 7, 2000\\
\vspace{.3in} ABSTRACT
\end{center}

We show that the Husain-Kucha\v{r} model can be described in the
framework of BF theories. This is a first step towards its
quantization by standard perturbative QFT techniques or the
spin-foam formalism introduced in the space-time description of
General Relativity and other diff-invariant theories. The actions
that we will consider are similar to the ones describing the
BF-Yang-Mills model and some mass generating mechanisms for gauge
fields. We will also discuss the role of diffeomorphisms in the
new formulations that we propose.

\vspace*{1cm} \noindent PACS number(s): 04.20.Cv, 04.20.Fy


\pagebreak

\setcounter{page}{1}

\section{Introduction}

BF theories were introduced by Horowitz \cite{Hor} and,
independently, by Blau and Thompson \cite{BlTh} in the late
eighties and early nineties. They are simple diff-invariant models
that  have been extensively studied as a testbed for quantization
techniques in the absence of metric backgrounds. They have also
been used as building blocks for physical theories such as gravity
\cite{Smo1} and Yang-Mills \cite{Mart}. From the point of view of
their quantum treatment it is interesting to point out that they
can be quantized both by using traditional Quantum Field Theory
(QFT) techniques and methods specially tailored to deal with
diff-invariant background-free actions such as spin networks and
spin-foams \cite{baez}.

The Husain-Kucha\v{r} model (HK) \cite{HK} is an interesting
diff-invariant system with local degrees of freedom that actually
mimics many of the features of 3+1 dimensional general relativity
(GR). In fact, the solutions to the Einstein equations (with an
internal $SO$(3) group) can be seen as a subset of the solutions
to HK. Its Hamiltonian description is very similar to the Ashtekar
formulation for GR because both share the same phase space and
most of the constraints --only the Hamiltonian constraint is
missing in HK--. Its absence means that the model describes
equivalence classes of metrics\footnote{\noindent Under both
diffeomorphisms and $SO$(3) internal rotations.} in the spatial
slices of a 3+1 foliation of space-time, but no time evolution. It
is widely believed that a successful quantization of HK, and the
implementation of efficient computational techniques to deal with
it, would be important steps towards the quantization of full GR.

The standard action for the HK model \cite{HK} is very similar to
the self-dual one \cite{Sam} for GR because they essentially
differ only in the internal gauge group ($SO$(3,1) for GR and
$SO$(3) for HK\footnote{\noindent Actually the self-dual action in
the Euclidean case can be obtained from  the HK action by adding a
term to it as shown in \cite{fer2}.}). This explains the
similarities of their Hamiltonian descriptions. Some other action
principles for this model have also appeared in the literature
\cite{fer1}-\cite{fer3}. They provide different Hamiltonian
formulations of the theory and solve some apparent shortcomings of
the original formulation --such as the description of
non-degenerate 4-metrics and the possibility to couple ordinary
matter--. One of them (the two-connection formulation \cite{fer1})
will be used in this paper as the starting point to obtain the HK
model by coupling BF Lagrangians. The action that we give is very
similar in form  to the ones appearing in \cite{Mart} for
BF-Yang-Mills (BFYM) models, in \cite{lah} for a mass generating
mechanism for gauge fields, or in \cite{baez} to discuss the
problem of infrared divergencies in the spin-foam quantization of
the BF theory.

There are several reasons that lead us to believe that the actions
that we consider here are of interest. First of all they are
structurally very simple because they consist only of several BF
terms plus quadratic interactions. At variance with the known
formulations, they have quadratic terms that  may make it simpler
to use perturbative QFT techniques. Also, we want to emphasize
that we are not resorting to the usual (and rather trivial) trick
of taking a BF model and impose restrictions on the 2-form field B
with Lagrange multipliers because we want to preserve the
quadratic character of the action as much as possible. Second, BF
theories have also been quantized by using the so called
spin-foams that provide a picture of quantum space-time geometry.
As is shown in \cite{baez} one may need to add a ``cosmological
constant" term to remove apparent infrared divergencies that pop
up in the computation of transition amplitudes\footnote{The
functional integrals in \cite{baez} are performed with the naive
measure in field space without introducing the terms needed to
take care of the second class constraints present in the theory.}.
This term is very similar to the ones that we introduce here so we
hope that the strong similarity of our formulation with this one
will provide a completely different way to deal with the HK model.
Some of the issues that are not yet fully understood in the
spin-foam framework, such as the relationship between the presence
of local degrees of freedom  and triangularization independence
may be illuminated by using the actions that we give here.

An additional reason to consider different formulations for old
models is the following. As it has been discussed by some authors
\cite{DMY}, \cite{EW1}  the non-renormalizability of GR can be
traced back to the non-invertibility of the quadratic part of the
action after gauge fixing of its symmetries. In the case of the
Hilbert-Palatini action  there is, in fact, no quadratic term to
build field propagators. Something similar happens for the
previously known actions for the HK model because all of them lack
a quadratic part. A possible way out of this is to consider the
non-quadratic actions as interaction terms of theories modified by
the addition of quadratic kinetic terms and try to extract some
information from these modified models. For this approach to work
it must be possible to add these kinetic terms in a consistent
way, i.e. the fully interacting theory must be a consistent
deformation (in the sense of \cite{Hen}) of its quadratic part. As
it turns out it is actually impossible to modify the theory by
adding kinetic terms (both diff-invariant and in the presence of a
metric background) to have a consistent formulation \cite{edufer}
where the Hilbert-Palatini or the HK actions appear as
interactions. This is a consequence of the fact that these actions
are built in terms of 1-form fields, so it is interesting to study
whether new formulations in terms of different objects --such as
2-forms-- will allow us to successfully incorporate consistent
kinetic terms, or hopefully, define propagators.

A final issue that we will discuss is the role of diffeomorphisms
in HK actions. We will show that, as in the 2+1 gravity case (a BF
theory itself), one can actually write down non-diff invariant
actions leading to  the same field equations and gauge symmetries
(on shell) as those of the standard formulations.

The paper is organized as follows. After this introduction we
give, in section II, the BF description of the HK model
--following a brief discussion of some of  the other known
formulations-- and  the general mechanism at work here and in
other BF descriptions of known theories. Section III contains our
 conclusions and comments. We end the paper with an appendix
where we discuss the detailed Dirac analysis for the actions that
we present in the paper.

\section{BF Formulation of the Husain-Kucha\v{r} Model}

BF theories in four dimensions are described by the action
\begin{equation}
S_{B\!F}=\int_{{\cal M}}d^4 x \;{\tilde\eta}^{abcd}B_{abi}
F_{cd}^i\,,  \label{001}
\end{equation}
where ${\tilde\eta}^{abcd}$ is the four-dimensional Levi-Civita
tensor density, $F_{ab}^{i}$ is the curvature of a gauge
connection 1-form $A_a^i$ (taking values in the Lie algebra of a
certain gauge group $G$) defined as
$F_{ab}^i=2\partial_{[a}A_{b]}^i+\left[A_a, A_b\right]^i$, and
$B_{abi}$ is a 2-form defined in the dual Lie algebra. Here and in
the following we use tangent space indices $a$, $b$, $c$,$\ldots$
In this paper we will restrict ourselves to $G=SO(3)$; and use
internal indices $i,\;j,\;k,...=1,2,3$. The totally antisymmetric
Levi-Civita tensor will be denoted as $\epsilon_{ijk}$. The
differentiable manifold ${\cal M}$ is taken to have the topology
${\cal M}=\real\times \Sigma$ with $\Sigma$ a three-dimensional
compact manifold without boundary.

The action for the Husain-Kucha\v{r} model is
\begin{equation}
S_{H\!K}=\int_{{\cal M}}d^4 x
\;{\tilde\eta}^{abcd}\epsilon_{ijk}e_a^i e_b^j F_{cd}^{k}.
\label{002}
\end{equation}
where $F_{ab}^i=2\partial_{[a}A_{b]}^i+\epsilon^{ijk}A_{aj}A_{bk}$
is the curvature of a $SO(3)$ connection $A_a^i$ and $e_a^i$ are
three 1-forms (notice that they are not proper tetrads because
there are only three of them). As shown in \cite{fer1} it can be
rewritten (modulo surface terms) as
\begin{equation}
S^{\prime}_{H\!K}=\int_{{\cal M}}d^4 x
\;{\tilde\eta}^{abcd}\fu_{abi} \fd_{cd}^{i}\, , \label{003}
\end{equation}
where $\fu_{abi}$ and $\fd_{abi}$ are the curvatures of two
$SO(3)$ connections
$\fu_{ab}^i=2\partial_{[a}\au_{b]}^i+\epsilon^{ijk}\au_{aj}\au_{bk}$
and
$\fd_{ab}^i=2\partial_{[a}\ad_{b]}^i+\epsilon^{ijk}\ad_{aj}\ad_{bk}$.
This can be easily seen by defining $e_{ai}=\ad_{ai}-\au_{ai}$ and
using the fact that
\begin{equation}
\fd_{ab}^i=\fu_{ab}^i+2\cu_{[a}e_{b]}^i+\epsilon^{ijk}e_{aj}
e_{bk} \label{004}
\end{equation}
(the covariant derivative $\cu_a$ acts on internal $SO(3)$ indices
as $\cu_a \lambda_i=\partial_a
\lambda_i+\epsilon_{i}^{\;\;jk}\au_{aj}\lambda_k$ and $\cd_a$ is
defined in an analogous way). By using (\ref{004}) we can write
the action (\ref{002}) as
\begin{equation}
S^{\prime}_{H\!K}=\int_{{\cal M}}d^4 x
\;{\tilde\eta}^{abcd}\left[\fu_{abi} \fu_{cd}^{i}+2\fu_{ab}^i\cu_c
e_{di}+\epsilon^{ijk}e_{ai}e_{bj}\fu_{cdk}\right];\label{005}
\end{equation}
dropping the first term due to its topological character and the
second after integrating by parts and using the Bianchi identities
we end up with the action (\ref{002}). Notice that both
$\fu_{ab}^i$ and $\fd_{ab}^i$ are defined in the Lie algebra of
$SO(3)$ so we must use the invariant metric $\delta ^{ij}$ to
build the action. The field equations coming from (\ref{003}) have
the beautifully symmetric form
\begin{equation}
\left\{
\begin{array}{l}
\cu_{[a} \fd_{bc]}^i=0 \\
\hspace{1cm} \\
\cd_{[a} \fu_{bc]}^i=0\, .
\end{array}\right.
\label{006}
\end{equation}
Let us consider now the following action\footnote{Here we are
following  a suggestion due to Giorgio Immirzi to couple two
Yang-Mills theories by using their BFYM formulations.}
\begin{equation}
S_{{\scriptscriptstyle B\!F\!H\!K}}=\int_{{\cal M}}d^4 x
\;{\tilde\eta}^{abcd}\left[\bu_{abi} \fu_{cd}^{i}+\bd_{abi}
\fd_{cd}^{i}+\bu_{abi}\bd_{cd}^{i}\right] \label{007}\,,
\end{equation}
obtained by coupling two BF theories via a
${\tilde\eta}^{abcd}\bu_{abi}\bd_{cd}^{i}$ interaction term. The
field equations coming from (\ref{007}) are now
\begin{equation}
\left\{
\begin{array}{l}
\fu_{ab}^i+\bd_{ab}^i=0\\
\hspace{1cm} \\
\fd_{ab}^i+\bu_{ab}^i=0
\end{array}\right.\hspace{1cm}
\left\{
\begin{array}{l}
\cu_{[a}\bu_{bc]}^i=0\\
\hspace{1cm} \\
\cd_{[a}\bd_{bc]}^i=0\,.
\end{array}\right.\label{008}
\end{equation}
The equivalence between these equations and (\ref{006}) is
obvious. The new action $S_{{\scriptscriptstyle \!B\!F\!H\!K}}$ is
both $SO(3)$ and diff-invariant. To understand how we get local
degrees of freedom from the two topological BF models it suffices
to realize that an action obtained by adding two different BF
terms has two independent sets of symmetries that are partially
broken by the ${\tilde\eta}^{abcd}\bu_{abi}\bd_{cd}^{i}$ term;
specifically, without this coupling the action would be invariant
under two independent sets of $SO(3)$ rotations and
diffeomorphisms. The reduced gauge symmetry accounts for the new
local degrees of freedom described by (\ref{007}). By using the
terminology introduced by Smolin in \cite{Smo1} the action
(\ref{007}) describes the HK model as a constrained topological
field theory.

There is an evident, one to one, correspondence between the
solution spaces to equations (\ref{006}) and (\ref{008}) so,
classically, both actions (\ref{003}) and (\ref{007}) describe the
same theory. The equivalence of their quantum formulations is,
however, less obvious. As $S_{{\scriptscriptstyle \!B\!F\!H\!K}}$
is quadratic in $\bu_{ab}^i$ and $\bd_{ab}^{i}$ one would be
tempted to say that the effect of performing the functional
integration in these fields would be equivalent to substituting
the field equations for them in the action, in which case we would
end up with the 2-connection action (\ref{003}) and a functional
integral measure involving only the connections. However this is
not correct because the functional measure is not the naive one
but must be modified \cite{HT} to take into account the existence
of second class constraints in the Hamiltonian formulation derived
from (\ref{007}). An intuitively clear way to understand this fact
is to realize that, as a consequence of the Bianchi identities,
only those components of $\bu_{ab}^i$ and $\bd_{ab}^{i}$ that
cannot be written as covariant derivatives (in terms of the
connections $\au_{ab}^i$ and $\ad_{ab}^{i}$) couple to the
curvatures $\fu_{ab}^i$ and $\fd_{ab}^{i}$ respectively. One must,
in fact, remove them form the path integral in order to avoid
problems such as the violation of the Ward identities that one
encounters in BFYM models \cite{Mart}. After modifying the measure
to circumvent this difficulty the integration in $\bu_{ab}^i$ and
$\bd_{ab}^{i}$ is no longer trivial. The Dirac analysis of
(\ref{007}) is rather involved because one expects to find many
second class constraints. Most of them appear as consistency
conditions for the solvability of the Lagrange multiplier
equations that one must introduce in the total Hamiltonian. A way
to partially alleviate this problem is to use St\"{u}ckelberg's
procedure \cite{Stu} in (\ref{007}) by introducing auxiliary
fields $\etu_a^i$ and $\etd_a^i$ in the action and trading some
second class constraints for first class constraints. So we will
also consider
\begin{equation}
S^{\prime}_{{\scriptscriptstyle B\!F\!H\!K}}=\int_{{\cal M}}d^4 x
\;{\tilde\eta}^{abcd}\left[\bu_{abi} \fu_{cd}^{i}+\bd_{abi}
\fd_{cd}^{i}+(\bu_{abi}-\cu_a
\etu_b^i)(\bd_{cd}^{i}-\cd_c\etd_d^i)\right] \label{009}\,.
\end{equation}
The field equations derived from (\ref{009}) are
\begin{equation}
\begin{array}{l}
                 \left\{
                        \begin{array}{l}
                  \fu_{ab}^i+\bd_{ab}^i-\cd_{[a}\etd_{b]}^i=0\\
                \hspace{1cm}\\
                  \fd_{ab}^i+\bu_{ab}^i-\cu_{[a}\etu_{b]}^i=0\, ,
                \end{array}\right.\\
\hspace{1cm}\\
            \left\{
                \begin{array}{l}
2\cu_{[a}\bu_{bc]}^i-\epsilon^{ijk}
\etu_{[a}^j\left(\bd_{bc]}^k-\cd_b\etd_{c]}^k\right)=0\\
\hspace{1cm}\\
2\cd_{[a}\bd_{bc]}^i-\epsilon^{ijk}
\etd_{[a}^j\left(\bu_{bc]}^k-\cu_b\etu_{c]}^k\right)=0\, ,
                \end{array}\right.
\end{array}
\label{010}
\end{equation}
$$
            \left\{
                \begin{array}{l}
\cu_{[a}\left(\bd_{bc]}^i-\cd_b \etd_{c]}^i\right)=0\\
\hspace{1cm}\\
\cd_{[a}\left(\bu_{bc]}^i-\cu_b \etu_{c]}^i\right)=0\,.
                \end{array}\right.
$$
From the first pair we find
\begin{equation}
\left\{
\begin{array}{l}
\bu_{ab}^i=-\fd_{ab}^i+\cu_{[a}\etu_{b]}^i\\
\hspace{1cm}\\
\bd_{ab}^i=-\fu_{ab}^i+\cd_{[a}\etd_{b]}^i\, ,
\end{array}
\right. \label{011}
\end{equation}
so that the remaining equations give (\ref{006}). After solving
them we can obtain $\bu_{ab}^i$ and $\bd_{ab}^i$ (modulo the
arbitrariness in $\etu_a^i$ and $\etd_a^i$) from (\ref{011}).
Notice that the third pair of equations in (\ref{010}) is just a
consequence of the first and the Bianchi identities.

In addition to the $SO(3)$ and the diff-invariance of (\ref{007})
this action is invariant under the transformations
\begin{equation}
\left\{
\begin{array}{l}
\delta \au_a^i=0\\ \\
\delta \ad_a^i=0\\ \\
\delta\bu_{ab}^i=\cu_{[a}\fiu_{b]}^i\\ \\
\delta \bd_{ab}^i=0\\ \\
\delta\etu_a^i=\fiu_a^i\\ \\
\delta \etd_a^i=0
\end{array}
\right. \quad \quad\quad\quad\quad\quad
 \left\{
\begin{array}{l}
\delta \au_a^i=0\\ \\
\delta \ad_a^i=0\\ \\
\delta \bu_{ab}^i=0\\ \\
\delta\bd_{ab}^i=\cd_{[a}\fid_{b]}^i\\ \\
\delta \etu_a^i=0\\ \\
\delta\etd_a^i=\fid_a^i
\end{array}
\right. \label{012}
\end{equation}
where $\fiu_a^i$ and $\fid_a^i$ are arbitrary gauge parameters. A
possible way to ensure that these are the only symmetries of the
proposed action (and a first step towards the quantization of the
model) is to study its Hamiltonian formulation. This is done in
the appendix. One can see that the model is, in fact, equivalent
to the 2-connection formulation after performing a suitable gauge
fixing of the symmetries introduced by the St\"{u}ckelberg
procedure. Here we follow a different path to prove the
equivalence of $S_{{\scriptscriptstyle B\!F\!H\!K}}$ and
$S^{\prime}_{{\scriptscriptstyle B\!F\!H\!K}}$ with (\ref{003})
{\it at the quantum level} by using the covariant symplectic
techniques introduced in \cite{CR1}.

To this end we must look at the symplectic form in the space of
solutions to the field equations \cite{CR1}. This is obtained in a
two-step process; first we write a closed (but degenerate at this
stage) 2-form in the space of fields. For a Lagrangian of the type
$L(\varpi, d\varpi)$ depending on a $s$-form field $\varpi$ and
its exterior derivative it is given by
\begin{equation}
\Omega=\int_{\Sigma}\dd \varpi \w \dd \frac{\partial L}{\partial d
\varpi}\,, \label{013}
\end{equation}
where we define $\frac{\partial L}{\partial d \varpi}$ according
to
$$
L(d\varpi+d\epsilon)=L(d\varpi)+d\epsilon \wedge \frac{\partial
L}{\partial d \varpi}+{\rm higher\;\;order}\,.$$
 Here we use the
differential form notation to make a clear distinction between
space-time and phase space objects. We must distinguish between
the ordinary exterior differential in the spacetime manifold
${\cal M}$ ($d$) and the exterior differential in the field
space($\dd$). In the same way we must make a distinction between
the wedge product in both cases ($\wedge$ and $\w$ respectively).
In the following we will avoid this by explicitly writing tangent
space indices. Notice that $d$ is defined in ${\cal M}$ but the
integral in (\ref{013}) is three-dimensional.

We must now pull-back $\Omega$ to the space of solutions to the
field equations. Whenever this can be explicitly done (in those
few cases where the solutions to the field equations can be
completely parametrized \cite{edufer}) one gets a closed and
non-degenerate 2-form in the space of solutions to the field
equations. The non-degeneracy implies that no gauge freedom
remains\footnote{This is equivalent to the reduced phase space of
the usual Hamiltonian formulation.} and the objects that appear in
the symplectic form parametrize physical degrees of freedom. For
the action (\ref{007}) we have\footnote{Notice that the $A(x)$ and
$B(x)$ fields are time dependent; here $d^3x$ denotes the measure
in $\Sigma$.}
$$
\Omega=2\int_{\Sigma}d^3x\,{\tilde \eta}^{abc}\left[\dd \au_a^i(x)
\w \dd \bu_{bci}(x) + \dd \ad_a^i(x) \w \dd \bd_{bci}(x)\right]\,,
$$
where ${\tilde \eta}^{abc}$ is the three dimensional Levi-Civita
tensor density in the $x$-coordinate patch  and $\au_a^i, \ad_a^i,
\bu_{ab}^i$, and $\bd_{ab}^i$ are the pull-backs of the
corresponding four dimensional objects onto $\Sigma$. After
partially pulling back to the solution space by using the field
equations expressing $\bu$ and $\bd$ in terms of the curvatures we
get\footnote{This coincides with the result in \cite{fer1}. This
example illustrates the efficiency of covariant symplectic methods
as compared to the more traditional approach of Dirac.}
\begin{equation}
\Omega=-4 \int_{\Sigma}d^3x\,{\tilde \eta}^{abc}\left\{\dd
\au_a^i(x) \w [\cd_b-\cu_b] \dd\ad_{ci}(x)\right\}. \label{015}
\end{equation}
This is not yet the symplectic structure in the solution space
because we still have to consider the remaining field equations
but it coincides with the one derived from (\ref{003}). This not
only proves the equivalence of the classical theories but also of
their quantum formulations because their reduced phase spaces
coincide and have the same symplectic structure. For the action
(\ref{009}) we obtain in a similar fashion
\begin{eqnarray}
\Omega\!=\!\int_{\Sigma}d^3x\;{\tilde \eta}^{abc}\left\{2\dd
\au_a^i(x) \!\w \dd \bu_{bci}(x) + 2\dd \ad_a^i(x) \!\w \dd
\bd_{bci}(x)-\right. \hspace{2cm}\label{016}
\end{eqnarray}
$$ \hspace{2cm}\left.-\dd\etu_a^i(x)\!\w\dd
[\bd_{bci}(x)-\cd_b\etd_{ci}(x)]-\dd\etd_a^i(x)\!\w\dd
[\bu_{bci}(x)-\cu_b\etu_{ci}(x)]\right\}$$ and, as before, after
partially pulling back to the solution space by using the field
equations expressing $\bu$ and $\bd$ in terms of curvatures we get
\begin{equation}
\Omega\!=\!\int_{\Sigma}\!d^3x{\tilde \eta}^{abc} \left\{\dd
\etu_a^i(x)\!\w\dd\fu_{bci}(x)+\!\dd
\etd_a^i(x)\!\w\dd\fd_{bci}(x)+ \right.\hspace{3cm} \label{017}
\end{equation}
$$
\hspace{1cm}\left.+2\dd \au_a^i(x) \!\w \dd [\cu_b\etu_{ci}(x)-
\fd_{bci}(x)]+\!2\dd \ad_a^i(x) \!\w \dd
[\cd_b\etd_{ci}(x)-\fu_{bci}(x)]\right\}
$$
which, after a little algebra gives precisely (\ref{015}) and,
hence, we also prove the classical and quantum equivalence of
(\ref{009}) with the actions considered before.

It is interesting to compare this situation with some other
similar models. Let us consider first the BFYM theory that is
known to be amenable to perturbative treatment \cite{Mart}. The
BFYM action can be written  as
\begin{eqnarray*}
S_{{\scriptscriptstyle B\! F\! Y\! M}}[A,B]=S_{Y\!
M}[A]-\int_{\cal M} d^4 x\;
(B_{ab}^i+\frac{1}{g}F_{ab}^i)(B_{cdi}+\frac{1}{g}F_{cdi})\eta^{ac}\eta^{bd}\,,
\end{eqnarray*}
where $S_{Y\! M}[A]$ is the usual Yang-Mills action for $A$ with a
coupling constant $g$ and $\eta^{ab}$ is a background Minkowski
metric. The term $(B+F/g)^2$ vanishes on shell thus making the
BFYM and the YM models equivalent\footnote{This argument, based on
the resolution of algebraic field equations back in the action,
relies on the following fact:  $S_{\scriptscriptstyle B\! F\! Y\!
M}$ action generates a covariant phase space that is in one-one
correspondence with the Yang-Mills one. Moreover, it induces a
symplectic two form over the space of $A$, $B$ fields that
coincides, when restricted  to the solution space, with the
symplectic Yang-Mills form.}. This term yields a propagator for
$B$. Moreover $S_{Y\! M}$ has, after a suitable gauge fixing, a
well defined propagator for the Yang-Mills field $A$.

\noindent{I}n the BFHK case we can also write
\begin{eqnarray}
S_{{\scriptscriptstyle B\! F\! H\! K}}[A_1,A_2,B] = \int_{\cal
M}d^4
x{\tilde\eta}^{abcd}(\bd_{ab}^i+\fu_{ab}^i)(\bu_{cdi}+\fd_{cdi})
-S^{\prime}_{H\! K}[A_1,A_2]\,,\label{a}
\end{eqnarray}
but now $S^{\prime}_{H\! K}$ does not have a free part and the
quadratic form $(\bd+\fu)(\bu+\fd)$ is always degenerate  in the
$A$, $B$ fields, even after gauge fixing. This prevents us from
using standard QFT techniques because we do not have a suitable
propagator even though we have a quadratic piece in the action.
However (\ref{a}) suggests a possible way out. The main   problem
that we face is related to  the diff-invariance of the quadratic
terms derived from $S_{{\scriptscriptstyle B\! F\! H\! K}}$.
Looking at the mechanism that is working in (\ref{a}), one can
devise a straightforward generalization where the diff-invariance
is broken through the use of a metric background $g$, so let us
consider, for example, the action
\begin{eqnarray}
S^{\prime\prime}=S^{\prime}_{H\! K}-\int_{\cal M}
(\bu_{ab}^i+*\fu_{ab}^i)
(*\bd_{cdi}+\fd_{cdi})\eta^{ac}\eta^{bd}\,.\label{b}
\end{eqnarray}
where $*F_{ab}^i$ is the Hodge dual $F_{ab}^i$. Once again, the
equivalence with the HK model is due to the vanishing of the
$B$-term of the action in the covariant phase space. Clearly,
although $S^{\prime\prime}$ is not diff-invariant, the
diff-invariance is recovered on shell.  This formulation has the
same perturbative objections that $S_{{\scriptscriptstyle B\! F\!
H\! K}}$. However, the presence of a metric background opens up
the possibility of adding to the action $S^{\prime\prime}$ kinetic
terms for $A$ and $B$. Then $S^{\prime\prime}$ might be considered
as an interaction term of an action with well-defined propagators.
Of course this would be possible only if these (local) kinetic
terms can be incorporated in a consistent way, both from the
classical and quantum point of view.

In addition to the BFYM models some other theories of this type
have been studied in the literature. An interesting example is the
mass generating mechanism for gauge fields proposed by Lahiri in
\cite{lah} and described by the action
\begin{equation}
\int_{{\cal M}}d^4x\,Tr\left[-\frac{1}{6}
H_{abc}H^{abc}-\frac{1}{4}F_{ab}F^{ab}+\frac{m}{2}{\tilde
\eta}^{abcd} B_{ab} F_{cd}\right]\,, \label{c}
\end{equation}
with $H_{abc}=3\nabla_{[a}B_{bc]}$ and $m$ is a mass parameter.
The essential difference between our model and Lahiri's is related
to how the $B$ field appears in the action. In \cite{lah}, after
disregarding total derivatives, one can write for the linear
theory
\begin{eqnarray}
\int_{{\cal M}}d^4x Tr\left\{-\partial_{[a}A_{b]}\partial^a
A^b-\frac{m^2}{4}A_a A^a
-\frac{3}{2}[\partial_{[a}B_{bc]}-\frac{m}{6}\epsilon_{abcd}
A^d][\partial^a
B^{bc}-\frac{m}{6}\epsilon^{abce}A_e]\right\}\label{free}
\end{eqnarray}
but now the $(dB-m*A)^2$ term does not vanish  on shell but it is
proportional to the gradient of the gauge parameter for the $A$
field. Another property of \cite{lah}, that makes his model closer
to $S^{\prime\prime}$, is the impossibility of using standard
perturbative treatments \cite{Henneaux}. This is so because the
free Lagrangian (\ref{free}) has two independent gauge symmetries
\begin{eqnarray*}
\begin{array}{lccl}
\delta_1 A_a^i=\partial_a\Lambda^i &\quad\quad& \quad\quad& \delta_2A_a^i=0\\
\delta_1 B_{ab}^i=0 & \quad\quad& \quad\quad& \delta_2
B_{ab}^i=\partial_{[a}\Lambda_{b]}^i\,,
\end{array}
\end{eqnarray*}
for arbitrary 0-forms $\Lambda^i$ and 1-forms $\Lambda_a^i$ while
the full model has only a $SO(3)$ symmetry. This problem is
exactly the same that afflicts the Palatini gravitational theories
in four space-time  dimensions \cite{edufer}. The open question
is, again, the  existence of consistent theories, in the
perturbative sense, for interacting 1-form and 2-form fields
implementing this type mechanism. As a systematic study of
quadratic actions for 1-form and 2-form fields has not been
carried out to date we hope that suitable theories can be found.

\section{Conclusions and Comments}

As we have shown in the paper the Husain-Kucha\v{r} model can be
obtained by coupling BF theories. The mechanism at work in the
actions (\ref{007}) and (\ref{009}) is different from the standard
procedure of writing a BF action and imposing constraints on the
2-form field with Lagrange multipliers. In the examples that we
consider in this paper the local degrees of freedom of the model
(three per space point) appear due to the breaking of some of the
symmetries of the BF terms induced by the introduction of a
``cosmologial constant" term coupling the 2-form fields $\bu$ and
$\bd$. The main reason why we do this is that we want to have a
well defined quadratic part in the action in order to define
propagators (after a suitable gauge fixing of the symmetries) and
treat the model by using standard perturbative QFT techniques.

Of course, even in the presence of a quadratic part, the theory
may not have suitable propagators. This happens if the symmetries
of the quadratic part do not match the symetries of the full
action (i.e. the full action is not a consistent deformation of
its quadratic part). In the present case we do not get a
perturbatively quantizable action because, as shown in
\cite{edufer}, its kinetic term is diff-invariant so it describes
no local degrees of freedom and hence it has more symmetries than
the full action. Anyway, it may be possible to modify these
actions by the addition of local kinetic terms that may render the
quadratic part invertible in such a way that the full action is a
consistent deformation of it. If this can be achieved it may be
possible to effectively recover the HK model in some limit, or at
least extract useful information from its behavior as an
interaction term. One of the points that we emphasize in the paper
is precisely the fact that writing the action in terms of new
types of fields it may be possible to have kinetic terms not
available in the known formulations.

We have shown that one can actually derive the HK models from non
diff-invariant actions. This is similar to GR because, as is well
known \cite{LL}, it is possible to derive the Einstein equations
from a non-diff-invariant action, quadratic in the Christoffel
connections, obtained by integrating by parts the second order
terms in the expression for the scalar curvature. In this case the
loss of symmetry is compensated by the appearance of many second
class constraints that reduce the number of physical degrees of
freedom to the two physical ones. Diff-invariance is kept only
on-shell. Again, the availability of a background structure may
permit the introduction of kinetic terms that could not be written
without its help.

We want to point out that covariant symplectic techniques are of
great help to study the physical content of field theories without
having to rely on the more traditional Dirac analysis. We have
shown that it is very easy to prove the classical and quantum
equivalence of the actions that we have discussed here. This
requires some comments. In principle, two non-trivially different
actions\footnote{By non-trivial we mean that they not differ in
total derivatives or cannot be transformed into one another by
simple redefinitions of the objects in terms of which they are
described.} may lead to the very same field equations. This does
not mean, however, that they are equivalent from the quantum point
of view. To prove their quantum equivalence the actions must endow
their solution spaces with the same symplectic structure. This
does indeed happen in this case as shown above but this should not
happen in general.

In a precise sense it is not necessary to go through the painful
procedure of using perturbative techniques as in \cite{Mart} to
show that the BFYM action is equivalent, from the quantum point of
view to the usual Yang-Mills one. Perturbation theory, if properly
done (and disregarding difficult issues of convergence), {\sl
reduces to canonical quantization}. If one can argue, as we did
before, that the solution spaces of the usual Yang-Mills
formulation and BFYM coincide and are endowed by the action with
the same sympletic structure their quantum formulations must
coincide (which is the result that one finds out, at the end of
the day).

\section{Appendix: Dirac Analysis of the Coupled BF Action for the HK Model}

We give a detailed Dirac analysis for the action (\ref{009}) to
show that we arrive exactly at the same conclusions that we got by
using covariant symplectic techniques. We start by introducing a
foliation of the space-time manifold ${\cal M}=\real\times \Sigma$
(where $\Sigma$ denotes a compact 3-manifold without a boundary)
defined by a scalar field $t$ and a congruence of curves nowhere
tangent to the $t=constant$ hypersurfaces parametrized by $t$. In
this way, if $t$ is used as a ``time coordinate'' we have
$\partial_t={\cal L}_{{\bf t}}$, i.e. time derivatives are given
by the Lie derivatives along the tangent vectors $\bf t$ to the
curves of the congruence parametrized by $t$. In the following we
make no distinction between three dimensional and four dimensional
tangent indices as it will be clear from the context  if we are
considering 3-dim or 4-dim objects. By using the identity ${\tilde
\eta}^{abcd}=4t^{[a}{\tilde \eta}^{bcd]}$, where ${\tilde
\eta}^{bcd}$ is the Levi-Civita tensor density in three
dimensions, and defining
\begin{equation}
\begin{array}{lll}
\left\{
       \begin{array}{l}
             t^a \bu_{ab}^{\;\;\;\;i}\equiv\but_b^i\\
             \hspace{1cm}\\
             t^a \bd_{ab}^{\;\;\;\;i}\equiv\bdt_b^i
       \end{array}
\right. & \left\{
       \begin{array}{l}
             {\tilde \eta}^{abc} \fu_{bci}\equiv\tfu_i^a\\
             \hspace{1cm}\\
             {\tilde \eta}^{abc} \fd_{bci}\equiv\tfd_i^a
       \end{array}
\right. & \left\{
       \begin{array}{l}
             t^a\etu_a^i\equiv\etu_{{\scriptscriptstyle 0}}^i\\
             \hspace{1cm}\\
             t^a\etd_a^i\equiv\etd_{{\scriptscriptstyle 0}}^i
       \end{array}
\right. \\
\hspace{1cm} & \hspace{1cm} & \hspace{1cm}\\
\left\{
       \begin{array}{l}
             {\tilde \eta}^{abc} \bu_{bc}^{\;\;\;\;i}\equiv\tbu_i^a\\
             \hspace{1cm}\\
             {\tilde \eta}^{abc} \bd_{bc}^{\;\;\;\;i}\equiv\tbd_i^a
       \end{array}
\right. & \left\{
       \begin{array}{l}
             t^a\au_a^i\equiv\au_{{\scriptscriptstyle 0}}^i\\
             \hspace{1cm}\\
             t^a\ad_a^i\equiv\ad_{{\scriptscriptstyle 0}}^i
       \end{array}
\right. & \hspace{1cm}
\end{array}
\label{A1}
\end{equation}
we find
\begin{eqnarray}
& & \int d t \left\{\!\tbu^a_i{\cal L}_{\bf
t}\!\au^i_a+\tbd^a_i{\cal L}_{\bf t}\!\ad^i_a+\frac{1}{2}\left(
{\tilde \eta}^{abc}\cd_b\etd^i_c-\tbd^a_i\right){\cal L}_{\bf
t}\etu^i_a+\frac{1}{2}\left(
{\tilde \eta}^{abc}\cu_b\etu^i_c-\tbu^a_i\right){\cal L}_{\bf t}\etd^i_a+\right.\nonumber\\
& & \hspace{1cm}\nonumber\\
& & +\but_a^i\left(\tfu^a_i+\tbd^a_i-{\tilde
\eta}^{abc}\cd_b\etd^i_c\right)+
\bdt_a^i\left(\tfd^a_i+\tbu^a_i-{\tilde \eta}^{abc}\cu_b\etu^i_c\right)+\label{A2}\\
& & \hspace{1cm}\nonumber\\
& & +\au^i_{{\scriptscriptstyle
0}}\left[\cu_a\tbu^a_i+\frac{1}{2}\epsilon^{ijk}\left(\tbd^a_j-
{\tilde \eta}^{abc}\cd_b\etd^j_c\right)\etu_{ak}\right]+
\frac{1}{2}\etu^i_{{\scriptscriptstyle 0}}\cu_a\left({\tilde
\eta}^{abc}\cd_b\etd^i_c-
\tbd^a_i\right)+\nonumber\\
& & \left.+\ad^i_{{\scriptscriptstyle
0}}\left[\cd_a\tbd^a_i+\frac{1}{2}\epsilon^{ijk}\left(\tbu^a_j-
{\tilde
\eta}^{abc}\cu_b\etu^j_c\right)\etd_{ak}\right]+\frac{1}{2}\etd^i_{{\scriptscriptstyle
0}}\cd_a\left({\tilde
\eta}^{abc}\cu_b\etu^i_c-\tbu^a_i\right)\right\}.\nonumber
\end{eqnarray}
The  momenta canonically conjugate to the
variables\footnote{Functional derivatives with respect to ${\cal
L}_{\bf t}$ of the corresponding fields.} $\au_a^i$, $\ad_a^i$,
$\tbu^a_i$, $\tbd^a_i$, $\au_{{\scriptscriptstyle 0}}^i$,
$\ad_{{\scriptscriptstyle 0}}^i$, $\but^a_i$, $\bdt^a_i$,
$\etu^a_i$, $\etd^a_i$, $\etu_{{\scriptscriptstyle 0}}^i$, and
$\etd_{{\scriptscriptstyle 0}}^i$ will be denoted as $\piu^a_i$,
$\pid^a_i$, $\sigu^i_a$, $\sigd^i_a$, $\piu_i$, $\pid_i$,
$\nuu^a_i$, $\nud^a_i$, $\psiu^a_i$, $\psid^a_i$, $\psiu_i$, and
$\psid_i$. They satisfy the following primary constraints
\begin{equation}
\begin{array}{lll}
\left\{
       \begin{array}{l}
             \piu_i^a-\tbu^a_i=0\\
             \hspace{1cm}\\
             \pid_i^a-\tbd^a_i=0
       \end{array}
\right. &

\left\{
       \begin{array}{l}
            \piu_i=0\\
            \hspace{1cm}\\
            \pid_i=0
       \end{array}
\right. &

\left\{
       \begin{array}{l}
             \sigu_a^i=0\\
             \hspace{1cm}\\
             \sigd_a^i=0
       \end{array}
\right. \\
\hspace{1cm} & \hspace{1cm} & \hspace{1cm}\\
\left\{
       \begin{array}{l}
             \nuu^a_i=0\\
             \hspace{1cm}\\
             \nud^a_i=0
       \end{array}
\right. &

\left\{
       \begin{array}{l}
             \psiu_i=0\\
             \hspace{1cm}\\
             \psid_i=0
       \end{array}
\right. & \left\{
       \begin{array}{l}
             \psiu^a_i-\frac{1}{2}\left({\tilde \eta}^{abc}\cd_b\etd_c^i-\bdt^a_i\right)=0\\
             \hspace{1cm}\\
             \psid^a_i-\frac{1}{2}\left({\tilde
             \eta}^{abc}\cu_b\etu_c^i-\but^a_i\right)=0\,.
       \end{array}
\right.
\end{array}
\label{A3}
\end{equation}
The Hamiltonian is
\begin{eqnarray}
& & H=-\int_{\Sigma}\!d^3 \!y
\left\{\but_a^i\left[\tfu^a_i+\tbd^a_i-{\tilde \eta}^{abc}\cd_b
\etd_{ci}\right]+\bdt^i_a\left[\tfd^a_i+\tbu^a_i-{\tilde
\eta}^{abc}\cu_b \etu_{ci}\right]+
\right.\nonumber\\
& & \hspace{1cm}\nonumber\\
& & +\au_{{\scriptscriptstyle 0}}^i\left[\cu_a\tbu^a_i+\frac{1}{2}
\left(\tbd^a_j-{\tilde \eta}^{abc}\cd_b\etd_{cj}\right)
\etu_{ak}\right]+\label{A4}\\
& & \hspace{1cm}\nonumber\\
& & \hspace{5cm}+\ad_{{\scriptscriptstyle
0}}^i\left[\cd_a\tbd^a_i+ \frac{1}{2}\epsilon_{ijk}\left(\tbu^a_j-
{\tilde \eta}^{abc}\cu_b\etu_{cj}\right)\etd_{ak}\right]+\nonumber\\
& & \left.+\frac{1}{2}\etu^i_{{\scriptscriptstyle 0}}
\cu_a\left[{\tilde\eta}^{abc}\cd_b\etd_{ci}-\tbd^a_i\right]+
\frac{1}{2}\epsilon_{ijk}\etd^i_{{\scriptscriptstyle 0}}
\cd_a\left[{\tilde\eta}^{abc}\cu_b\etu_{ci}-\tbu^a_i\right]\right\}\nonumber
\end{eqnarray}
and the total Hamiltonian
$$ H_T\!=\!H\!+\!\int_{\Sigma}\!d^3\!y\!\left\{\lu_a^i(\piu^a_i\!-\!\tbu^a_i)
\!+\!\ld_a^i(\pid^a_i\!-\!\tbd^a_i)\!+\!
\lu_i\piu^i\!+\!\ld_i\pid^i\!+\!\tmu^a_i \sigu_a^i\!+\!\tmd^a_i
\sigd_a^i\!+\!\su_i\psiu^i\!+\!\sd_i\psid^i\!\!+\!\right.$$
\vspace*{-.5cm}
\begin{equation}
\left.\!+\!\nnu_a^i\nuu^a_i\!+\!\nnd_a^i\nud^a_i\!+\!
\su_a^i\left[\psiu^a_i\!-\!\frac{1}{2}\left({\tilde\eta}^{abc}\cd_b
\etd_c^i\!-\!\tbd^a_i\right)\right]\!+\!
\sd_a^i\left[\psid^a_i\!-\!\frac{1}{2}\left({\tilde\eta}^{abc}\cu_b\etu_c^i\!-
\!\tbu^a_i\right)\right]\right\}\,,\label{A5}
\end{equation}
where $\lu_a^i$, $\ld_a^i$, $\lu_i$, $\ld_i$, $\tmu^a_i$,
$\tmd^a_i$, $\nnu_a^i$, $\nnd_a^i$, $\su_a^i$, $\sd_a^i$, $\su_i$,
and $\sd_i$ are Lagrange multipliers. By imposing stability of the
primary constraints under the evolution defined by the total
Hamiltonian $H_T$ we get secondary constraints and equations
involving the Lagrange multipliers. The secondary constraints are
\begin{equation}
\begin{array}{lll}
\!\!\!\!\!\!\left\{\!\!\!
       \begin{array}{l}
             \cu_a\tbu^a_i\!-\!\epsilon_{ijk}\psiu^a_j \etu_{ak}\!=\!0\\
             \hspace{1cm}\\
             \cd_a\tbd^a_i\!-\!\epsilon_{ijk}\psid^a_j \etd_{ak}\!=\!0
       \end{array}
\right. & \!\!\!\!\left\{\!\!\!
       \begin{array}{l}
             \tfu^a_i\!+\!\tbd^a_i\!-\!{\tilde\eta}^{abc}\cd_b\etd_{ci}\!=\!0\\
             \hspace{1cm}\\
             \tfd^a_i\!+\!\tbu^a_i\!-\!{\tilde\eta}^{abc}\cu_b\etu_{ci}\!=\!0
       \end{array}
\right. &
 \!\!\!\!\left\{\!\!\!
       \begin{array}{l}
             \cu_a\left[{\tilde\eta}^{abc}\cd_b\etd_{ci}\!-\!\tbd^a_i\right]\!=\!0\\
             \hspace{1cm}\\
             \cd_a\left[{\tilde\eta}^{abc}\cu_b\etu_{ci}\!-\!\tbu^a_i\right]\!=\!0.
       \end{array}
\right.
\end{array}
\label{A6}
\end{equation}
We also find a set of equations that the Lagrange multipliers must
satisfy. In this case the stability of the secondary constraints
does not give any new ones but only additional equations for the
Lagrange multipliers. These equations must be treated with care.
In many cases it is possible to show that they can be solved for
any configuration of the fields in which case it is not necessary
to solve them explicitly in order to get the final form of the
constraints. However, it may happen\footnote{As, for example, if
one considers the action (\ref{007}).} that some restrictions on
the fields and momenta must be imposed in order to guarantee their
solvability. These conditions are new secondary constraints that
must be put in equal footing with the remaining ones, in
particular we must impose their stability that may lead to new
equations and constraints. In the present case a long and tedious
computation shows that no new constraints must be enforced to
ensure the solvability of the equations for the Lagrange
multipliers.

By using the explicit expression for the Lagrange multipliers and
the fact that, once they are plugged in $H_T$ we can express it as
the sum of a first class Hamiltonian and a linear combination of
the primary first class constraints with arbitrary functions of
time as coefficients we find out that $\au_{{\scriptscriptstyle
0}}^i$, $\ad_{{\scriptscriptstyle 0}}^i$,
$\etu_{{\scriptscriptstyle 0}}^i$, $\etd_{{\scriptscriptstyle
0}}^i$, $\but_{{\scriptscriptstyle 0}}^i$,
$\bdt_{{\scriptscriptstyle 0}}^i$ are arbitrary functions and the
following first class constraints
\begin{eqnarray}
& & \tfu^a_i+\tbd^a_i-{\tilde\eta}^{abc}\cd_b\etd_{ci}=0\nonumber\\
& & \tfd^a_i+\tbu^a_i-{\tilde\eta}^{abc}\cu_b\etu_{ci}=0\nonumber\\
& & 2\cu_a\tbu^a_i+\epsilon_{ijk}\left(\tbd^a_j-{\tilde\eta}^{abc}
\cd_b\etd_{cj}\right)\etu_{ak}=0\label{A7}\\
& & 2\cd_a\tbd^a_i+\epsilon_{ijk}\left(\tbu^a_j-
{\tilde\eta}^{abc}\cu_b\etu_{cj}\right)\etd_{ak}=0\nonumber
\end{eqnarray}
where the last two equations can be substituted by
\begin{eqnarray}
\cu_a\tfd^a_i=0\nonumber\\
\cd_a\tfu^a_i=0\label{A8}
\end{eqnarray}
by using the first pair. We also find the following second class
constraints
\begin{eqnarray}
& & \piu^a_i-\tbu^a_i=0\nonumber\\
& & \pid^a_i-\tbd^a_i=0\nonumber\\
& & \sigu_a^i=0\nonumber\\
& & \sigd_a^i=0\label{A9}\\
& & 2\psiu^a_i-\left({\tilde\eta}^{abc}\cd_b\etd_c^i-\tbd^a_i\right)=0\nonumber\\
& &
2\psid^a_i-\left({\tilde\eta}^{abc}\cu_b\etu_c^i-\tbu^a_i\right)=0\,.\nonumber
\end{eqnarray}
The symplectic structure for the initial set of canonical
variables (after the elimination of the arbitrary
objects$\au_{{\scriptscriptstyle 0}}^i$, $\ad_{{\scriptscriptstyle
0}}^i$, $\etu_{{\scriptscriptstyle 0}}^i$,
$\etd_{{\scriptscriptstyle 0}}^i$, $\but_{{\scriptscriptstyle
0}}^i$, $\bdt_{{\scriptscriptstyle 0}}^i$) is
$$\Omega=-2\int_{\Sigma}d^3 x\left[\dd\au_a^i(x)\w\dd\piu^a_i(x)+
\dd
\tbu^a_i(x)\w\dd\sigu_a^i(x)+\dd\etu_a^i(x)\w\dd\psiu^a_i(x)+\right.$$
$$\hspace{3cm}\left.+\dd\ad_a^i(x)\w\dd\pid^a_i(x)+
\dd
\tbd^a_i(x)\w\dd\sigd_a^i(x)+\dd\etd_a^i(x)\w\dd\psid^a_i(x)\right]\,.$$
Now we must pull it back onto the hypersurface in phase space
defined by the second class constraints. As a first step we use
(\ref{A9})  to remove $\sigu_a^i$, $\sigd_a^i$, $\tbu^a_i$, and
$\tbd^a_i$. At this stage we are left with the first class
constraints
\begin{eqnarray}
& & \tfu^a_i+\pid^a_i-{\tilde\eta}^{abc}\cd_b\etd_{ci}=0\nonumber\\
& & \tfd^a_i+\piu^a_i-{\tilde\eta}^{abc}\cu_b\etu_{ci}=0\nonumber\\
& & \cu_a\tfd^a_i=0\label{A11}\\
& & \cd_a\tfu^a_i=0\,,\nonumber
\end{eqnarray}
the second class constraints
\begin{eqnarray}
& & 2\psiu^a_i-\left({\tilde\eta}^{abc}\cd_b\etd_c^i-\pid^a_i\right)=0\nonumber\\
& &
2\psid^a_i-\left({\tilde\eta}^{abc}\cu_b\etu_c^i-\piu^a_i\right)=0\,,\label{A12}
\end{eqnarray}
and the symplectic structure
$$\Omega=-2\int_{\Sigma}d^3 x\left[\dd\au_a^i(x)\w\dd\piu^a_i(x)+
\dd\etu_a^i(x)\w\dd\psiu^a_i(x)+\right.\hspace{3cm}$$
$$\hspace{7cm}\left.+\dd\ad_a^i(x)\w\dd\pid^a_i(x)+
\dd\etd_a^i(x)\w\dd\psid^a_i(x)\right].$$ We have now two possible
choices:

\noindent i) Remove $\psiu^a_i$ and $\psid^a_i$ by using
(\ref{A12}) to get a phase space coordinatized by $\au_a^i$,
$\ad_a^i$, $\piu_a^i$, $\pid_a^i$, $\etu_a^i$, and $\etd_a^i$, the
first class constraints (\ref{A11}) and the symplectic structure
$$\Omega=-\int_{\Sigma}d^3 x\left[2\dd\au_a^i(x)\w\dd\piu^a_i(x)+
\dd\etu_a^i(x)\w\dd
\left({\tilde\eta}^{abc}\cd_b\etd_{ci}(x)-\pid^a_i(x)\right)+\right.$$
$$\hspace{3cm}\left.+2\dd\ad_a^i(x)\w\dd\pid^a_i(x)+
\dd\etd_a^i(x)\w\dd
\left({\tilde\eta}^{abc}\cu_b\etu_{ci}(x)-\piu^a_i(x)\right)\right]\,.$$
As we have 54 canonical variables and 24 first class constraints
we have three physical degrees of freedom per space point. The
Dirac brackets of the phase space variables can be obtained in
closed form by inverting the previous expression for $\Omega$ in
matrix form.

\noindent ii) Remove $\piu^a_i$ and $\pid^a_i$ to get a phase
space coordinatized by $\au_a^i$, $\ad_a^i$, $\psiu_a^i$,
$\psid_a^i$, $\etu_a^i$, and $\etd_a^i$. The first class
constraints are given now by
\begin{eqnarray}
& & \tfu^a_i-2\psiu^a_i=0\nonumber\\
& & \tfd^a_i-2\psid^a_i=0\nonumber\\
& & \cu_a\tfd^a_i=0\nonumber\\
& & \cd_a\tfu^a_i=0\nonumber
\end{eqnarray}
and the symplectic structure is
$$\Omega=-2\int_{\Sigma}d^3 x\left[\dd\au_a^i(x)\w\dd
\left({\tilde\eta}^{abc}\cu_{b}\etu_{ci}(x)-
2\psid^a_i(x)\right)+\dd\etu_a^i(x)\w\dd\psiu^a_i(x)+\right.$$
$$\hspace{3cm}\left.+\dd\ad_a^i(x)\w\dd\left({\tilde\eta}^{abc}\cd_{b}\etd_{ci}(x)-
2\psiu^a_i(x)\right)+\dd\etd_a^i(x)\w\dd\psid^a_i(x) \right].$$ As
before we have 54 canonical variables and 24 first class
constraints per space point. In both cases we can recover in a
straightforward way the formulation of \cite{fer1} by using the
consistent gauge fixing conditions
$$\etu_a^i=0$$
$$\etd_a^i=0.$$
If instead of using the action (\ref{009}) we take (\ref{007}) the
procedure is considerably more involved due to the appearance of
several layers of secondary constraints that come up as
consistency conditions for the solvability of the Lagrange
multipliers equations that are avoided here by using the
St\"{u}ckelberg procedure. Anyway, as argued in the main text, the
formulation is strictly equivalent to the other ones presented in
the paper.

\end{document}